\documentstyle[12pt,aaspp4]{article}

\def\sun{\hbox{$\odot$}}

\lefthead{Majewski et al.}
\righthead{Sagittarius dSph at $b=-40\deg$}

\begin{document}
\title{Starcounts Redivivus: III. A Possible Detection of the Sagittarius Dwarf 
Spheroidal Galaxy at $b=-40^o$}

\author{S. R. Majewski\altaffilmark{1,2,3}, M. H. Siegel\altaffilmark{1}, 
\\W.~E. Kunkel\altaffilmark{4}
\\I.~N. Reid\altaffilmark{6},
\\K.~V. Johnston\altaffilmark{5},
\\I.~B. Thompson\altaffilmark{7},
\\A.~U. Landolt\altaffilmark{2,8},
\\C. Palma\altaffilmark{1}
}

\altaffiltext{1}{Dept. of Astronomy, University of Virginia,
Charlottesville, VA, 22903-0818 (srm4n@didjeridu.astro.virginia.edu, 
mhs4p@virginia.edu, cp4v@virginia.edu)}

\altaffiltext{2}{Visiting Research Associate, The Observatories of the Carnegie
Institution of Washington, 813 Santa Barbara Street, Pasadena, CA 91101}

\altaffiltext{3}{David and Lucile Packard Foundation Fellow, Cottrell Scholar
of The Research Corporation}

\altaffiltext{4}{Las Campanas Observatory, Carnegie Institution of
Washington, Casilla 601, La Serena, Chile
(skunk@roses.ctio.noao.edu)}

\altaffiltext{5}{Institute for Advanced Study, Olden Lane,
Princeton, NJ 08540}

\altaffiltext{6}{California Institute of Technology, 105-24, Pasadena, CA 91125 
(inr@astro.Caltech.Edu)}

\altaffiltext{7}{The Observatories of the Carnegie Institution of Washington,
813 Santa Barbara Street, Pasadena, CA 91101 (ian@ociw.edu)}

\altaffiltext{8}{Dept. of Physics \& Astronomy, Louisiana State University, 
Baton Rouge, LA 70803-4001 (landolt@rouge.phys.lsu.edu)}

\begin{abstract}

As part of the Selected Areas Starcount Survey (SASS), a CCD survey to 
$V>21$,  
we have obtained $VI$ photometry of two fields at $b=\pm 40^{\circ}$
aligned roughly with an extrapolation of the major 
axis of the Sagittarius (Sgr) dwarf spheroidal galaxy.  
Comparison of the color-magnitude diagram (CMD) for some of these
fields to the CMDs of fields reflected about the Galactic $l=0^{\circ}$ meridian
reveals an excess of stars at $V_o=17.85$ and $0.9 < (V-I)_o < 1.1$ in the 
$(l,b) = (11^{\circ},-40^{\circ})$ field.  The excess stars have colors consistent with 
the Sgr red clump, and deeper CMD imaging in these locations shows evidence
of a main-sequence turnoff at $V=21$ with the main-sequence 
extending to the limit of our data ($V=24$).  The surface brightnesses we derive
from either the potential excess of red clump stars or the apparent excess of 
MSTO stars are consistent with each other and with the results of other
surveys at this latitude.
No similar excess appears in our northern Galactic hemisphere fields near
$l=353^{\circ}, b=+41^{\circ}$ field. 

We have obtained spectroscopy of all 30 candidate red clump
stars in the range $0.9 < (V-I)_o < 1.1$ and $ 17.75 < V_o < 17.95 $.   
The radial velocity distribution of the stars, while dissimilar from 
expectations of Galactic structure models,
does not show a contribution by stars near the Galactocentric radial velocity
seen in other studies near the Sgr core.  It is difficult to
reconcile a photometric result that is consistent with other explorations of the
Sagittarius stream with a radial velocity distribution that is apparently 
inconsistent.
In a companion paper 
(Johnston et al. 1999), we discuss how some of the
discrepancies are resolved if {\it our} potential Sgr detection corresponds
to a {\it different} Sgr tidal streamer than that detected by 
most other surveys.

\end{abstract}


\clearpage
\section {Introduction}

The discovery in 1994 (Ibata, Gilmore \& Irwin 1994, I94) of the 
Sagittarius (Sgr) dwarf spheroidal (dSph) galaxy has provided direct 
observational support for the proposition that dwarf satellite galaxies 
can be tidally disrupted by the gravitational potential of the 
Milky Way and contribute to both its field star and cluster populations.  
Stars associated with Sgr but well beyond its tidal radius of $\sim4^{\circ}$ 
(Ibata et al. 1997, I97 hereafter) have been 
detected by a variety of techniques, some that we utilize in the present study.
Deep color magnitude diagrams (CMDs) of fields oriented along the major axis of 
Sgr have revealed the presence of the Sgr main sequence turn off (MSTO)
superimposed on the CMD of
field stars (Mateo et al. 1996, hereafter M96; Fahlman et al. 1996, F96; 
Mateo et al. 1998, M98).  
Surveys of RR Lyrae stars have shown concentrations at the distance modulus of 
Sgr (M96; Alard 1996, A96; Alcock et al. 1997, A97).  The prominent Sgr 
red clump has also proven useful for delineating its structure (I94; Sarajedini
\& Layden 1995, SL95; Ibata et al. 1995, I95).  These stellar Sgr detections are 
plotted in Aitoff projection in Figure 1 along with planetary nebulae
(Zijlstra \& Walsh 1996) and globular 
clusters associated with Sgr (Terzan 7, Terzan 8, Arp 2 and M54).  
Figure 1 demonstrates the growing extent of the Sgr ``tidal stream" as 
the joint surveys probe a larger area. 

We have begun the Selected Areas Starcounts Survey (SASS) to
provide accurate photometric catalogues for our continuing study
of the classical Galactic starcounts problem (Reid \& Majewski 1993).  
Among our 
completed SASS fields, Kapteyn Selected Areas (SA) 107 ($l=6^{\circ}, 
b=41^{\circ}$) and SA184 ($356^{\circ}, -40^{\circ}$) were imaged 
in the {\it BVRI} passbands with an approximately square, 5x5 grid of 
overlapping CCD images covering 2.1 deg$^2$ per SA, of which a subset 
(1.26 deg$^2$) has been analyzed in the context of Galactic starcounts 
already (Siegel et al. 1997, SMRT97 hereafter).  To constrain the 
symmetry of the halo and Intermediate Population II thick disk near 
$l = 0^{\circ}$, 
we also observed ``anti-fields" (ASA107 and ASA184 in Figure 1), reflected 
about the Galactic $l=0^{\circ}$ meridian, in $V$ and $I$.
Because these ``anti-fields" 
lay near the projection of the Sgr major axis, however, we 
rearranged our normal square grid of individual CCD images to a linear strip 
optimized to detect the Sgr stream, if present at these latitudes.  In this way, 
we might evaluate better the extent to which our starcount analysis might be 
sensitive to, and affected by, interception of Sgr-like tidal structures.

We outline here the identification of possible interloping Sgr stars in 
our ASA184 field, at $b=-40^{\circ}$, as first suggested by Siegel et al.
(1997).  In that report, we presented results based on excesses in the
field star color magnitude diagram that were intriguing, unique
in our data set, and consistent with the presence of Sgr in the one
survey area most likely to intercept the Sgr debris stream, though, admittedly,
the photometric evidence was not decisive.   Since that report, however,
additional support for a Sgr interpretation to the photometric excesses we
found has come from the work by M98 who have obtained deep CMDs 
in fields near our ASA184 field and who claim unambiguous detection 
of Sgr stars.  The M98 results (particularly the measured surface brightness of Sgr
debris in this part of the sky) are remarkably consistent with our own
results.   

Though our findings here may be argued to remain non-decisive, there are 
several reasons that motivate making our results available: (1)
Our data would seem to corroborate (with great consistency) the 
independent work of M98 to trace the dwindling (and more difficult to 
follow) density profile of the Sgr debris stream at great angular 
separations from the Sgr core. (2)  There is great interest in, indeed
need for, tracing out the full extent of the Sgr debris stream; however,
because of the huge area covered by Sgr, mapping its extent will require
a huge investment by the community.  In some sense, ``every little bit
helps" (note the number of different groups contributing 
to the Sgr mapping in Figure 1), and our findings can help advance
this endeavour by providing an additional benchmark in one region
of the sky. (3)  We have supplemented our photometry with spectroscopy of 
the supposed Sgr red clump stars in the ASA184 field and have produced a radial 
velocity distribution.  Thus, if the Sgr interpretation is born out, 
we provide the only velocity measure of the debris stream at such 
large distances from the Sgr core.  (4) While the velocity distribution 
we derive does show a curious trend that is suggestive of 
the presence of a ``moving group", the mean radial velocity of the stars is 
unlike expectations extrapolated from previous radial velocity studies of Sgr
near its core.  This has inspired a new dynamical model, 
presented in a companion paper (Johnston et al. 1999, ``Paper II" hereafter), 
of the destruction and present state of the Sgr dwarf spheroidal.
The new model, which is unique in that it accomodates all previous Sgr observations 
(as well as our new and unexpected results) naturally and self-consistently,
explains the low latitude results (those presented here and by M98) as 
being derived from a separate, but overlapping
Sgr tidal tail stripped off on a previous perigalactic passage of Sgr.
(5) There is value in pointing out anomalous findings in field star
studies, even if only to point out phenomena meriting further study, 
to alert colleagues of potential problem areas,
and to underscore our present lack of understanding 
of the nature of the Galactic stellar populations.  We point
to the precedent of Rodgers et al. 1990, whose presentation of an 
unusual excess population in a similar direction of the sky bears
some resemblance to our report here.

\section {Photometric Observations and Analysis}

The fields ASA184 and ASA107 were observed in the $V$ and $I$ passbands
as part of the SASS program in July, 1996 on the Swope 1-meter 
telescope at Las Campanas using the thinned $2048^{2}$ TEK5 CCD chip.  The 
large format CCD at the 29."03 mm$^{-1}$ Swope image scale gives a 
23.'5 field of view.  In each of these two antifields, a narrow strip of
images was taken approximately perpendicular to the major axis of Sgr, 
with field centers displaced by 20' 
to retain some overlap as a check on the photometry.  However, some 
fields are displaced laterally as much as 20' from this alignment so as  
to avoid bright foreground stars (see Figure 2).

Reductions were carried out with the IRAF package CCDRED.  Photometry 
was performed with the IRAF version of DAOPHOT (Stetson 1987) using a variable 
PSF.  Photometric calibration was obtained via Landolt (1992) standards, the 
IRAF package PHOTCAL, and our own code.  Photometry presented here reaches a 
depth of $V=21$ in nine ASA184 subfields (1.18 deg$^2$),
eight ASA107 subfields (1.07 deg$^2$), nineteen SA107 
subfields (2.42 deg$^2$) and fifteen SA184 subfields (1.96 deg$^2$).  
Photometric errors at $V=18.5$ are ($\sigma_{V},\sigma_{V-I}$) = (0.02,0.03).

The CMD's of ASA184 and ASA107 are shown in Figure 3.  In this paper, we 
adopt reddening coefficients, $E_{B-V}$, of 0.05 for 
ASA107, 0.07 for for SA107, 0.08 for ASA184 and 0.02 for SA184.  Differential
reddenings from CCD field to CCD field are estimated by intercomparing the peak of the 
blue edge on the ($V-I, V$) CMD in one magnitude brightness bins centered at 
$V=16.0, 16.5$ and $17.0$; the estimated error in the differential reddening 
from this technique is 0.01 magnitude.  
For all four fields, the relative reddening differences match the Burstein 
\& Heiles maps (1982) well, and, in the end, 
we set the absolute reddening from these maps.
We have taken $E_{V-I} = 1.24 E_{B-V}$ and $A_{V} = 3.1 E_{B-V}$ from 
Cardelli, Clayton \& Mathis (1989).

On first inspection, a distinctive feature in the CMD corresponding to
Sgr stars is not apparent in either ASA field in Figure 3 --  
any such signal is lost in the noise of the Galactic field star population.  
Any relative contribution by Sgr can be amplified by considering 
only stars with colors consistent with those expected for
the satellite's populous red clump and avoiding the prominent,
blue, MSTO edge of the field star CMD.
We therefore consider stars with $0.90< (V-I)_o <1.10$.
In this restricted color range, the field star magnitude distribution
rises monotonically to fainter magnitude, while the Sgr population 
should produce a sharp peak at the apparent magnitude appropriate
to the red clump.  I97 estimated 
the FWHM of this peak to be 0.1 magnitudes from a CMD in the center of the Sgr dSph.

Figure 4a shows a magnitude histogram of ASA184 
stars with $0.90<(V-I)_o<1.10$.  A slight peak is apparent in the 
ASA184 data at $V_o=17.85$, the magnitude expected for the red clump of Sgr
at a distance modulus of ($m-M$)=16.8 given the proper reddening correction 
and using the I97 absolute magnitude, $M_V$=1.04, for the red clump. 
The distance modulus would be close to the prediction of I97 for this latitude.
This $V_o=17.85$ overdensity is strongest in a limited range
of Galactic longitudes we have sampled.  The histogram for
stars in three
specific, adjacent ASA184 subfields (numbers 2-4 in Figure 2), 
centered at $(l,b) = (10.9,-40.2)$, 
or ($\alpha$, $\delta$)$_{2000}$ = (20:59:58, -33:21:01),
show that the contrast of this peak is increased from little more than 2$\sigma$
(Figure 4a) to more than 3$\sigma$ (Figure 4b).  
Figure 3c shows the CMD of these subfields with the SL95 Sgr isochrone 
superimposed, shifted to a distance modulus of 16.8.  
Unfortunately, the dominant field star population still allows no more than a
slight impression of a possible Sgr red clump in either Figures 3a and 3c.

We compare these results to the magnitude histograms in the same color range for 
ASA107, SA184 and SA107 (Figures 4c-4e, respectively).  No $V_o=17.85$ peak is seen in these
other fields, nor in comparison CMD regions -- given by $1.15 \le (V-I)_o < 
1.40$ -- in any of the fields including ASA184. 
Moreover, for either color range no field shows {\it any} 
other peak of nearly the same significance as the $V_o=17.85$ peak in ASA184, 
which suggests that this feature is unique in the data set.  The ASA184 peak is
robust to change in bin size and bin center, generally appearing at the 2-3 $\sigma$ level
of significance.

This statistical peak is not, in itself, strong evidence of the presence of Sgr.  
In a large survey parsed into small color-magnitude bins, one might expect to find occasional
deviations from a uniform distribution.  However, such a peak happening to fall 
near the magnitude and color of the expected Sgr clump at a latitude consistent 
with the expanding tidal stream of Sgr (and now consistent with the M98 detection) 
is suggestive and warrants further investigation.

If Sgr is present in these subfields of ASA184, it should be possible to 
identify its more populous upper main sequence and turnoff stars through deeper imaging.  
On UT 14-16 July 1997, we imaged two 0.022 deg$^{-2}$ ASA184 subfields using 
the du Pont 2.5-m and the same CCD used in the Swope observations.  One deep CCD
image was centered at the location of the possible red clump excess (ASA184 
subfield 3), the other approximately two degrees west along $b=-40^{\circ}$ (in
ASA184 subfield 8, which does {\it not} show the $V_o=17.85$ excess in the 
Swope data).  These new photometric data, which have a magnitude limit 
of $V_o=24.25$,  were reduced similarly to the shallower Swope data.  Figure 
5 shows the CMD of these two subfields.  An excess of objects at $(V-I)_o=0.8$ 
and $20.5 < V_o < 24$ appears in ASA184-3 (Figure 5a) compared to ASA184-8 
(Figure 5b).  In Figure 5c, the Figure 5a data are shown again with the 
F96 isochrone for Sgr plotted.  The excess in Figure 5a is matched by the
Sgr MSTO region if we assume, as before, a Sgr distance 
modulus of $(m-M)_o$ = 16.8

As further support for the
presence of a Sgr MSTO in ASA184-3, we show in Figure 5d the cumulative
starcounts over $17.75 < V_o < 22.25$ for ASA184-3 and ASA184-8 in two color 
ranges: $0.6 < (V-I)_o < 0.8$, which would bracket the main contribution of a 
Sgr MSTO and upper main sequence in the magnitude range shown, 
and $1.1 < (V-I)_o < 1.3$, which should miss most of a ($m-M)_o$ = 16.8, 
Sgr isochrone at these magnitudes.  We note that while the starcounts of the 
redder color range track each other well, the cumulative starcounts for the 
MSTO color range do indeed show a break near $V_o \approx 20.75$
in the ASA184-3 subfield which is not seen in the ASA184-8 subfield.  
A Kolmogorov-Smirnov test of the null hypothesis that the plotted cumulative
distributions for ASA184-3 and ASA184-8 are drawn from the same population
gives a signficance level of 0.95 for the redder color range (which yields
a total of 25 stars in ASA184-3 and 22 stars in SA184-8), but a level of 
only 0.23 for the MSTO color range (represented by 98 stars in ASA184-3 
and 67 stars in ASA184-8).  While this KS significance does not strongly rule 
out a similar parent population, 
0.23 is at the upper end of KS significances (as low as a few percent) 
found for various binnings tested 
(including wider bin widths with increased sample
sizes and better statistics that include more of the ``SGR MSTO" region)
blueward of $(V-I)_o = 0.95$.  
For the case illustrated in Figure 5d, the 
significance of the excess from Poissonian statistics is $2.7\sigma$.
Further deep imaging over more area in this
region of the sky, away from the ``spine" of fields covered by M98,
would be a great help to improving the statistics of these tests 
and verifying the width of the Sgr stream at these latitudes.

\section{Surface Density Considerations}

If we have indeed detected tidal stellar debris of Sgr, we can place some estimate
on the surface brightness of that debris.  The implied excess of stars in our red clump
CMD bin is some 10-14 stars in 0.41 deg$^2$ based on a comparison to our 
Galactic model (see below), which matches other parts of the CMD well.
Therefore, we adopt 35 red clump stars deg$^{-2}$ as an approximate upper limit
on the excess.

Note that the lowest isopleth plotted in I95 for red clump stars is 1800 deg$^{-2}$, and this is 
expected to be a $\sim 50\%$ completeness contour.  
Thus, the density of our red clump stars $26^{\circ}$ from the Sgr core at M54 
appears to be lower by about two orders of magnitude from the lowest I95 isopleths
which extend to about $5^{\circ}$ from the Sgr core near M54,
where the red clump densities reach a factor of roughly three higher
still (but note that the red clump density of SL95
stars 12 arcmin north of M54 is roughly 1800 deg$^{-2}$ -- see I97 Figure 2 --
a factor of two below the I95 and I97 isopleths at this point).  

In I97, Sgr maps are given utilizing stars within a magnitude of the MSTO.
In this case, isopleths of MSTO
stars at 3600 deg$^{-2}$ reach to about $13^{\circ}$ from the Sgr core at M54 
(where the Sgr MSTO density is a factor of 10 higher). From 
Figure 5d, we see that our excess of stars within one magnitude of the
MSTO is roughly 14 stars in .022 deg$^{2}$, and this yields a density of MSTO stars $26^{\circ}$
from the Sgr core of $\sim 1200$ deg$^{-2}$, a fall off of only a factor of three
across the $13^{\circ}$ from the last I97 point 
\footnote[1]{Because I97 do not explicitly name the color range 
used in their isopleth map, it is not clear whether
we have sampled exactly the same region of the Sgr MSTO.}.  
The latter suggests a more gradual falloff of Sgr light in the last 
$13^{\circ}$ along the stream to the point of our detection than in the 
first $13^{\circ}$ from the Sgr core south.  Such a ``break" in the 
surface density of Sgr is precisely what is observed by M98 in their
systematic survey along the length of the Sgr southern tidal arm.

In M98, the surface density of Sgr MSTO stars is derived to a 
Galactic latitude of $b=-48^{\circ}$ using a slightly different bin in ($I,V-I$) space, which
we estimate as $19.6<I<20.3$ and $0.4<V-I<0.8$.  At this 
latitude, M98 detects a density of approximately 1200 stars deg$^{-2}$, which
includes both background Galactic stars and Sgr MSTO stars in approximately equal
numbers.  Our data detect 26 stars in this region of color-magnitude space in the deep 
0.033 deg$^{2}$ observation of ASA184-3, 
which corresponds to 1200 stars deg$^{-2}$, a result identical to the M98 result.  This 
would place our surface brightness as identical to M98 at this 
latitude (29.8 mag arcsec$^{-2}$).

The surface brightness implied by the red clump stars should produce 
an identical result.  In I95, the density of Sgr red clump stars 
in the center of Sgr is set at 1.125 stars arc-minute$^{-2}$.  
Our corresponding density of 0.01 stars arc-minute$^{-2}$ (assuming 
only half of the stars to be Sgr as in M98) would place our surface 
brightness 5.0-5.7 magnitudes below the core surface brightness of Sgr, or 
30.2-30.9 mag arcsec$^{-2}$ -- a value slightly fainter, but 
fairly consistent with, the MSTO calculation.

\section{Radial Velocities}

In August 1997, September 1997, August 1998, and October 1998
we used the du Pont 2.5-m telescope at Las Campanas 
to obtain spectra of the stars in the $17.75 < V_o < 17.95$ 
peak of Figure 4b.  The celestial coordinates and undereddened colors and magnitudes
for these stars are listed in Table 1 (the dereddened colors and magnitudes used
in our analysis are obtained by assuming $E(V-I)=0.1$ and $A_V=0.25$).  
Typical errors in our $V$ and $V-I$ are about 0.015 and 0.020 magnitudes, 
respectively.  Our spectra cover the wavelength range 
from 4800-6800 \AA\ at a resolution of 2.5 \AA\ and include 
the Mgb triplet, MgH band (5211 \AA\  bandhead), and Na D doublets, 
which have been shown
(Deeming 1960, Friel 1987, Paltoglou \& Bell 1994, 
Tripicchio et al. 1997) to be useful discriminants between 
metal poor, evolved stars and more metal rich dwarfs.  The data also
encompass several iron lines useful for evaluating [Fe/H].  

The spectra are of sufficient signal-to-noise to allow measurement of line indices and
we anticipated being able to elucidate luminosity classes for our sample
of candidate Sgr red clump stars, segregating foreground Galactic dwarfs.
In fact, we found the results of such an analysis
\footnote[2]{Our analysis employed use of
(a) Friel's (1987) Mgb+H index, (b) a wider index 
sensitive to MgH absorption over 5000-5250 \AA\ 
(c) the equivalent width of the NaD doublet, fixed to 
continuum levels at 5875 and 5909 \AA\ , and (d) a variant
of the Friel (1987) [Fe/H] measure using five available iron lines, 
and for which Friel calibrated stars of luminosity class III and IV.}
to be ambiguous.  The separation of disk dwarfs from red clump stars -- which, for
the Mgb, MgH and NaD features, are expected to have line strengths 
intermediate between those of giants and dwarfs of
the same metallicity -- is complicated in the case of {\it Sgr} red 
clump stars by virtue of the apparently high mean metallicity 
(photometrically determined to be [Fe/H]=-0.52 by SL95) 
and large spreads of abundances in this galaxy.
As determined by the high resolution spectroscopic analysis
of Smecker-Hane, McWilliam \& Ibata (1998), Sgr has some metal-poor stars 
of [Fe/H] $\approx -1.5$; however this same analysis
indicates the existence of Sgr stars as metal rich as [Fe/H]=+0.11.
Marconi et al. (1999) report an even more extreme result in their
study of 23 candidate Sgr stars; these authors 
find {\it no} Sgr stars as metal poor as [Fe/H] = -1.5, and
apparently find a rather high mean abundance, and stars
as metal rich as [Fe/H]=+0.7.  The effect of such high abundances on the
line strengths of evolved stars is to counteract the suppression
of absorption associated with decreasing surface gravity.  
Contributing to this modulation of line strengths by [Fe/H]
is an apparent mean overabundance of magnesium in Sgr. From 
their table of representative Sgr stars, 
Marconi et al. (1999) find a mean [Mg/Fe] overabundance of +0.26.
Such ambiguities were possibly manifest in our analysis
\footnote[3]{The referee mentions having had similar problems in 
determining abundances for stars in the main body of Sgr, which he has 
found to have quite peculiar element ratios.}, which
yielded the bulk of the 30 stars in the sample to have line strengths
between those of metal rich giants and metal poor dwarfs, but, in many cases, 
with implied surface gravities dependent on which specific spectroscopic
indicator was selected.

Because of these and other difficulties, in the end 
we found a line strength analysis of our spectra to 
be premature, and beyond the scope of the present contribution.
Further work along these lines will require 
a larger sample of stars with improved signal-to-noise, a more
thorough calibration incorporating a sizable sample of {\it
bona fide} Sgr red clump stars for comparison, a consistent set of photometric
colors for both standard and target stars, and better stellar atmosphere 
models specifically constructed for red clump stars.  As a clearer
picture of the complex Sgr abundance patterns emerges,  it
may be possible to take better advantage of the line strength information
in our spectra.  We
can provide our spectroscopic data to interested parties upon request.

Nevertheless, our spectra are of suitable quality
for the measurement of radial velocities.
In Table 1 we present the heliocentric radial velocities derived for our red clump
sample.  Since all of 
our stars are in a small region of the sky, the offset from heliocentric 
to Galactocentric frames is nearly constant at +21 km s$^{-1}$, if 
a rotational velocity for 
the local standard of rest of 220 km s$^{-1}$ and a solar motion with 
respect to that of $(U,V,W) = (-9,11,6)$ km s$^{-1}$ are adopted.  
Velocities were obtained 
via a cross-correlation analysis against well established, bright
velocity standards.  To minimize flexure induced systematics, comparison 
arcs for each target and calibration source were always 
taken right after, and at the same telescope position as, the stellar spectra.
Strong spectral lines were left out of the cross-correlation analysis.
The location of the peak (``CCP") of the cross-correlation curve was taken as the 
radial velocity difference between target and calibration spectrum; 
the CCP value is included in Table 1.  The typical error in the radial velocities is 
12 km s$^{-1}$, as determined by repeat measures of a number of stars (including
numerous stars beyond those presented here) on different observing runs and 
after accounting for a mean run to run systematic offset velocity determined
from all repeat stars.  However, in some cases, due to low signal
to noise, a CCP was rather weak and the radial velocity 
is rather more suspect.  In cases where CCP $<0.30$ (radial velocities
marked as ``::" in Table 1) or $0.30 \le $CCP$ < 0.40$ (radial velocities
marked as ``:" in Table 1), an observation with 
higher signal to noise was sought.  For the subsequent analysis, we have 
adopted radial velocities with CCP $\ge 0.40$ over other measured radial
velocities for the same star if the latter velocities have CCP $< 0.40$. 
Otherwise, multiple radial velocity measures have been averaged.  

Because the field surveyed, ASA184, is almost directly below the Galactic 
center, the mean radial velocity of stars from both nearby and very distant
stars (whose mean rotational component is predominently perpendicular to 
the line of sight) 
are expected to be around $V_{helio} \lesssim $0 km s$^{-1}$.  For stars near
the Galactic center, the line of sight becomes tangent to their mean
rotation around the Galactic center and the radial velocity reflects 
more of their mean rotational velocity.  In Figure
6 we show the expected mean heliocentric radial velocities for stellar 
populations of different mean rotational rates around the Galactic center
as a function of distance from the Sun in the direction of the field ASA184.  
We consider four possible 
contaminants of our ``Sgr red clump" sample, which is centered in the CMD 
at $V_o=17.85$ and $(V-I)_o=1.0$:

\noindent{\it Halo giant stars:} At $V_o=17.85$ and $b=-40^{\circ}$ any giant
star with $(V-I)_o=1.0$ will be well into the halo.  If we adopt stars of
the metallicity of M3 and M13 as typical of the halo, we derive a 
distance of 37 kpc.  The expected mean radial velocity for such contaminants
would be $< -25$ km s$^{-1}$ (Figure 6) but with a rather large ($\sim 100$ km s$^{-1}$)
dispersion.

\noindent{\it Horizontal branch stars:} With $(V-I)_o=1.0$ these would 
most likely be red clump stars.  If we adopt the I97 absolute magnitude
for Sgr red clump stars as typical for ``contaminant" red clump stars
from the halo field, they would be at a distance of about 23 kpc.  The 
mean heliocentric radial velocity for these stars would be $< -10$
km s$^{-1}$, but again with a large velocity dispersion if from a random 
halo field population.

\noindent{\it Metal rich dwarfs:} If solar abundance, a $(V-I)_o=1.0$ 
dwarf would be 2.2 kpc distant and 1.4 kpc (some four old disk 
scaleheights) below the Galactic plane.  The expected mean
radial velocity for these stars would be $< 0$ km s$^{-1}$, and,
while of solar metallicity, the velocity dispersion would need 
to be of order 40 
km s$^{-1}$ or so for them to be at such large distances from the
Galactic mid-plane. 
One might not expect
an abundance of solar metallicity stars at this $z$-distance; 
we present them here as one limit for dwarf
contaminants, to compare to...

\noindent{\it Metal poor subdwarfs:} For a subdwarf 3 magnitudes below
the nominal solar abundance main sequence, the distance is some 550 pc, and
350 pc below the Galactic plane.  Thus, most dwarf contaminants in the sample
are likely to be somewhere between 0.5 and 2.2 kpc distant, and, from Figure
6, should show mean heliocentric velocities near $-10$ km s$^{-1}$ for 
typically expected, ``thick disk" asymmetric drifts.

What should we expect as ``contaminants" of our ``SGR red clump sample"
as predicted from standard Galactic models?
Using star count model B from Reid et al. (1996), which we have 
already shown to give reasonable fits to the SA107 and SA184 data 
(SMRT97; see also Figure 4), we predict 40-50 stars deg$^{-2}$ for
$17.75 < V_o < 17.95$ and $0.9<(V-I)_0<1.1$ at this $(l,b)$ (or 16-20 stars 
in our spectroscopic sample of 30 stars).  This expected number is in 
reasonable (though not exact) agreement with the density of stars (a) in 
this same color-magnitude bin for the Galactic structure fields ASA107, 
SA184 and SA107 (Figures 4c-e), and (b) with the density of stars in ASA184 
in the magnitude bins adjacent to the $V_o = 17.8$ bin in Figures 4a and 4b.  
In our model, about one third of the expected 16-20 stars 
are predicted to be from the disk and two-thirds from the thick disk;  
only a few stars 
are expected to be evolved ($M_V<3$) stars from a homogeneously distributed 
halo population (and these will be predominently giants). 
Assuming mean rotational velocities of $\sim 200$, $\sim 170$, 
and $\sim 40$ km s$^{-1}$ for the old thin disk, the thick disk, and the halo,
respectively, then we should predict some ten or so stars centered around
$V_{helio} \sim -10$ km s$^{-1}$ (Figure 6) with a dispersion of $\lesssim 50$ km s$^{-1}$
or so (mostly a combination of the $\sigma_{U}$ and $\sigma_{W}$ 
parts of the thick disk velocity ellipsoid), some five stars more tightly
clumped around $V_{helio} \sim -10$ km s$^{-1}$ (represeting the old thin disk),
and a few halo giants broadly spread around $V_{helio} \sim -30$ km s$^{-1}$.
It is not clear what population the ``excess" dozen or so stars should 
represent in the sample of 30 stars.

We find the mean heliocentric radial velocity for all 30 stars in the sample is 
4 km s$^{-1}$, with an RMS of 58 km s$^{-1}$.
At first blush, the distribution of velocities (Figure 7) 
seems {\it roughly} consistent with expectations for a mix of thin
and thick disk stars and a smattering of halo stars, but from Figure 6 and
the starcount model predictions detailed above
we should expect the velocities for any combination of stellar populations
to skew towards negative $V_{helio}$, rather
than positive $V_{helio}$ as is found (Figure 7).  We note, for example, that
16 of the 30 stars have $V_{helio} > 0$ (the median is 9.5  km s$^{-1}$), 
when the majority of stars should have $V_{helio} < 0$.  

Despite the skew to $V_{helio} > 0$, the lack in our data of a distinct radial 
velocity signal separate from 
that of the Galactic disk stars is unfortunate in that it does not allow us 
immediately to isolate specific members of any excess population above
the predicted Galactic stellar populations, and, indeed,
may even call into question the reality of such an excess population in the
spectroscopic sample.  However, we call attention to a {\it color dependence} 
of the radial velocity distribution which does seem to point to the presence of a
distinct radial velocity signal.  As shown in Figure 8, there is a dichotomy 
in the radial velocity distributions on either side of $(V-I)_o \sim 1.0$,
with a difference in the mean velocity of stars of 47 km s$^{-1}$.  The
bluer stars show a mean $V_{helio} = -17$ km s$^{-1}$ and velocity dispersion 
of 67 km s$^{-1}$ while the redder stars show a mean velocity of 
$V_{helio} = 30$ km s$^{-1}$ and much smaller velocity dispersion of 
only 27 km s$^{-1}$.  The velocity distribution {\it and} number of stars
in the bluer population are very similar to the expectations for 
Galactic stellar populations as outlined above.  

On the other hand,
the velocity dispersion for the redder stars is much less than expected for even
a random sample of only Galactic disk dwarf stars in this direction of the sky; 
however, the red star velocity dispersion is more
typical of expectations for halo ``moving groups", and we note that if this
redder sample should also contain some contribution by Galactic stars,
they would tend to {\it increase} the velocity dispersion.  Moreover, 
the {\it mean} radial velocity of the redder sample is more than 40 km s$^{-1}$ 
higher than expectations for Galactic disk stars. 
A Kolomogorov-Smirnov test that the radial velocity distributions on either
side of $(V-I)_o =1.0$ are drawn from the same population gives only 1\% 
confidence.  

The relevance of this velocity-color disparity may lie in a closer inspection of the
Sgr CMD in SL95 (see their Figures 4 and 6b) which seems to show (1) the Sgr red clump
concentrated to $1.0 <(V-I)_o < 1.15$ and (2) a relative
dearth of stars for $0.9 < (V-I)_o < 1.0$.  Perhaps a
more appropriate color bin for our selection of Sgr red clump stars should
have been $1.0 <(V-I)_o < 1.15$, and we should {\it expect} Sgr
red clump stars primarily in the $(V-I)_o > 1.0$ part of the spectroscopic
sample, but {\it not} the $(V-I)_o < 1.0$ part of the spectroscopic sample.  
While this would seemingly dovetail nicely with the color dependence 
of the radial velocities described above, we must also point out that
this parsing of the spectroscopic sample by color yields no consistent
{\it excess} in counts by our photometric data (both the red and blue sample show a 
similar low-significance excess).  That is, to identify the $1.0< (V-I)_o < 1.1$ stars
that show the much tighter velocity dispersion as a ``moving group" we 
have to believe simultaneously that the major contribution of the Galaxy
to our spectroscopic sample just happened to be in the $0.9< (V-I)_o < 1.0$
sample.  This is an unlikely assertion and stresses the need for a larger statistical 
sample to understand these apparent contradictions.

We conclude from this
analysis that {\it if} the Sgr red clump population is indeed present in our spectroscopic
sample, its mean velocity is likely to be rather low --  a mean velocity of 
$V_{helio} = 30$ km s$^{-1}$
or $V_{GSR} = 51$ km s$^{-1}$ if we adopt the $1.0< (V-I)_o < 1.1$ stars as
representative, and lower if we do not.  We note that most of the previous 
velocity surveys of Sgr near its core determined mean velocities greater than
$V_{GSR} = 150$ km s$^{-1}$, whereas we have only two stars with velocities that 
large.  As we discuss in the companion paper, this discrepancy with previous radial velocity 
work does not, in itself, discount our having detected the Sgr red clump
in our spectroscopic sample.

\section{Discussion}

\subsection{Sagittarius or Not Sagittarius?}

We are presented with the following set of apparently conflicting
circumstances: (1) the implication of our model and starcount data presented in 
Figure 4b is that there is a 50-100\% excess of stars in the ASA184
CMD in the magnitude and color range of interest; (2) this particular
part of the CMD is a reasonable match to the expected location of the prominent
Sgr red clump, and a Sgr-like isochrone and the appearance of a prominent
Sgr-like MSTO is implied by our deep CMD in this part of the sky, as 
presented in Figure 5; (3) the surface brightness of stars in these two regions 
are consistent with each other, consistent with an extrapolation of the Sgr 
extent into this field, and consistent with the results of M98; but (4) 
the radial velocity distribution of stars
in this field, while dissimilar from expectations for that of Galactic stars, 
is not grossly so when the  sample is considered in total; 
(5) there is a color trend in the velocity distribution, which is not supported by
the location of the count excess, but which does have an intriguing correspondence
to the color range of Sgr red clump stars in SL95; 
(6) we find the spectral line strength indices to yield ambiguous 
gravity/abundance information.

At this point, we are uncertain of how to account for (5) and both (5) and (6) 
clearly require further observations to resolve.  
We propose three ways to 
reconcile (1)-(4): (a) Our photometric results are merely a statistical fluke
in the large SASS project and these stars are all Galactic; 
(b) we have found a stellar system with an
extended spatial distribution that has a CMD extremely similar to Sgr,
is at about the same distance as Sgr, has about the same surface brightness
as Sgr in this part of the sky, but that is not Sgr;
or (c), we have indeed identified Sgr, but in our field Sgr has a 
very different radial velocity than seen near the center of the dwarf
galaxy.

Given the consistency of our photometric results with each other and with M98, 
we find proposition (a) unlikely, although not impossible.  A larger survey 
with deep imaging
would address this issue conclusively.  Proposition (b) would require the unique
circumstance of two distinct stellar systems overlapping in surface density and 
position, which is also unlikely, but not impossible.
In the companion paper 
(Johnston et al. 1999), we demonstrate how proposition (c) is viable
if our observations are revealing a {\it different} tidal tail of Sgr
than observed at higher Galactic latitudes.  
With semi-analytical modeling of the destruction of
the Sgr dwarf and by following the fate of debris shed by perigalacticon
passages of the dwarf on a variety of orbits, Johnston et al. conclude that 
lines of sight progressively farther from the Sgr core should intercept
additional debris streamers of Sgr with differing radial
velocities and distances.  
The possible detection of Sgr here,
with the distance and velocity characteristics we derive, match
remarkably well to the predictions for a tidal streamer torn from
Sgr {\it three} perigalactic passages ago.  Spectroscopy of Sgr
stars in the nearby M98 fields would be an important
test of the various propositions put forth here and in
Johnston et al. (1999).

\subsection{Sgr in the Northern Hemisphere}

  From the lack of a Sgr detection in our ASA107 field at $b=+40^{\circ}$, 
it is tempting to impose an {\it upper} limit to 
Sgr's possible extent in this direction.
However, we hedge on such an interpretation.  
In the first place, the orbital path of the dSph is not precisely known 
above the Galactic equator.  It is not only possible, but likely, 
that we missed the orbit entirely and that the ASA107 data do not constrain 
the length or age of the tidal stream.  It is worth noting that in a more 
careful
treatment of the Sgr disruption, Johnston (1998) shows that it would
take a mere 3 Gyr for Sgr tidal debris 
to wrap entirely around $360^{\circ}$, if it has a mass of $10^8 M_{\sun}$.

Most studies of the Sgr stream have concentrated on its 
southern extent since studying the northern tail 
means working into the Galactic plane, a difficult prospect.  
Nevertheless, observations at $b>10^{\circ}$ are possible and a Sgr component 
might be expected at $b \ge 10^{\circ}$ given the $25^{\circ}$ southern extent 
we may have detected in this paper and that is seen by M98.

\subsection{Implications for Galactic Structure Studies}

Finally, it is important to place our likely detection of Sgr within the
context of the original aims of our SASS survey, which were to constrain
global density models of Galactic stellar populations.  From the mere
existence of the Sgr system, it is evident that simple triaxial models
and symmetric analytic expressions
are insufficient to describe the outer populations of the Milky Way.  The
recent work by Johnston (1998) is particularly relevant to this discussion. 
She finds that after 10 Gyr, the fraction of the sky that would be covered
by the debris of a disrupted, $10^8 M_{\sun}$ Sgr would be on the order 
5000 deg$^2$, while the LMC should cover some 10,000 deg$^2$.  
However, to the
extent that there {\it do} exist more 
dynamically relaxed disributions of stars, 
the expansive diaspora of tidal debris may lurk as a mere tittle against an 
overwhelming stellar background (e.g., see Johnston 1998, Figure 8).  
With a wider canvas of SASS fields, we hope to evaluate the veracity
of simple analytical approaches to modeling the structure of the Milky 
Way, and determine the level of need 
for superposition of irregularities in model Galactic density distributions. 

\acknowledgements
 
This research was supported by NSF grants AST-9412265 to IT and SRM, 
AST-9412463 to INR, and AST-9528177 to AUL.  SRM also acknowledges 
support through a Fellowship from the David and Lucile Packard Foundation 
and a Cottrell Scholar Award from the Research Corporation.  We thank Rodrigo Ibata
for useful comments in the refereeing stage.
We dedicate this work to the memory of our friend and colleague, Jerry Kristian.

\clearpage
\begin{center}
TABLE 1.
\begin{tabular}{c|c|c|c|c|c|c|r@{.}l|c|r} \hline\hline
Star  & $\alpha$ & $\delta$ &$V$  &$V-I$ &  Obs Date   & Exptime & \multicolumn{2}{c|}{RV$_{helio}$} & CCP  \\
      & J2000.0  & J2000.0  &     &      &  mmddyy     & (sec)   & \multicolumn{2}{c|}{($km/sec$)}   &     \\
\hline
01 & 21:00:05.7 & -33:30:15 & 18.028 & 1.049 & 081197 &        2700 &  -137&5 ::&      0.25 \\
02 & 20:59:35.2 & -33:31:39 & 18.146 & 1.161 & 081197 &        1800 &   -10&8   &      0.45 \\
03 & 20:59:41.3 & -33:32:11 & 18.024 & 1.043 & 081198 &        900  &    99&2 : &      0.33 \\
   &            &           &        &       & 102898 &        1800 &   114&1   &      0.61 \\
04 & 21:01:04.8 & -33:34:30 & 18.199 & 1.075 & 102898 &        1440 &   -30&4   &      0.70 \\
05 & 21:01:33.7 & -33:39:55 & 18.115 & 1.073 & 081497 &        1350 &    10&6 ::&      0.22 \\
   &            &           &        &       & 081298 &        1800 &    19&6   &      0.86 \\
06 & 20:50:48.2 & -33:50:57 & 18.099 & 1.073 & 081298 &        1200 &  -113&6   &      0.75 \\
07 & 21:00:22.4 & -33:29:59 & 18.141 & 1.170 & 102898 &        1080 &    17&9   &      0.73 \\
08 & 20:59:26.2 & -33:33:17 & 18.022 & 1.007 & 082397 &        900  &   110&6 ::&      0.28 \\
   &            &           &        &       & 081298 &        1350 &   113&2   &      0.90 \\
09 & 20:59:58.6 & -33:35:31 & 18.027 & 1.034 & 081497 &        1350 &   -37&4 : &      0.37 \\
   &            &           &        &       & 081497 &        1350 &   -36&9 : &      0.37 \\
   &            &           &        &       & 081298 &        900  &   -51&0   &      0.70 \\
10 & 21:00:42.2 & -33:49:59 & 18.159 & 1.140 & 082497 &        900  &   -14&0   &      0.44 \\
11 & 20:59:26.2 & -33:49:47 & 18.023 & 1.140 & 081298 &        900  &    45&9   &      0.77 \\
12 & 21:00:07.5 & -33:09:56 & 18.066 & 1.057 & 081298 &        900  &    37&5   &      0.73 \\
13 & 21:00:12.7 & -33:20:34 & 18.130 & 1.188 & 081298 &        900  &    31&5   &      0.64 \\
14 & 21:01:01.5 & -33:23:30 & 18.065 & 1.135 & 082497 &        900  &    61&2 : &      0.30 \\
15 & 21:00:22.6 & -33:10:05 & 18.016 & 1.156 & 081298 &        900  &    67&1   &      0.59 \\
16 & 20:59:28.9 & -33:13:27 & 18.050 & 1.084 & 081397 &        1800 &    -3&5   &      0.51 \\
   &            &           &        &       & 081397 &        1800 &    -4&0   &      0.50 \\
17 & 20:59:57.6 & -33:14:41 & 18.177 & 1.004 & 081298 &        900  &  -145&9 : &      0.32 \\
   &            &           &        &       & 102898 &        1800 &   -73&7   &      0.63 \\
18 & 21:01:11.0 & -33:17:23 & 18.035 & 1.064 & 081298 &        3600 &   -15&5   &      0.72 \\
   &            &           &        &       & 102898 &        1800 &   -23&1   &      0.83 \\
19 & 21:01:17.1 & -33:21:56 & 18.018 & 1.164 & 081398 &        1440 &    22&7   &      0.77 \\
   &            &           &        &       & 102898 &        1800 &    -2&7   &      0.75 \\
\hline
\end{tabular}
\end{center}

\clearpage
\begin{center}
TABLE 1 (continued).
\begin{tabular}{c|c|c|c|c|c|c|r@{.}l|c|r} \hline\hline
Star  & $\alpha$ & $\delta$ & $V$ & $V-I$&  Obs Date   & Exptime & \multicolumn{2}{c|}{RV$_{helio}$} & CCP  \\
      & J2000.0  & J2000.0  &     &      &  mmddyy     & (sec)   & \multicolumn{2}{c|}{($km/sec$)}   &     \\
\hline
20 & 20:59:37.2 & -33:21:36 & 18.058 & 1.001 & 081297 &        1800 &  -34&9 ::&       0.20 \\
   &            &           &        &       & 102898 &        1800 &    9&1   &       0.58 \\
21 & 21:00:43.9 & -33:27:08 & 18.077 & 1.110 & 082397 &        900  &   65&3 : &       0.32 \\
   &            &           &        &       & 081398 &        1800 &   61&3   &       0.85 \\
24 & 21:01:15.3 & -33:33:56 & 18.146 & 1.189 & 102898 &        1800 &   46&9   &       0.60 \\
25 & 21:00:04.8 & -33:25:06 & 18.072 & 1.143 & 082397 &        900  &   10&2   &       0.45 \\
   &            &           &        &       & 081398 &        1080 &    7&4   &       0.73 \\
27 & 21:00:48.0 & -33:03:07 & 18.154 & 1.167 & 081398 &        2720 &   36&0   &       0.80 \\
28 & 20:59:38.8 & -33:06:46 & 18.081 & 1.027 & 081398 &        2720 &  -32&9 : &       0.35 \\
30 & 21:00:08.2 & -32:51:50 & 18.021 & 1.066 & 082497 &        900  &  -10&0 ::&       0.24 \\
   &            &           &        &       & 081398 &        1260 &  -51&5   &       0.69 \\
31 & 21:00:32.4 & -33:03:22 & 18.075 & 1.032 & 102898 &        1350 &  -37&4   &       0.47 \\
32 & 21:00:16.3 & -33:05:57 & 18.090 & 1.167 & 081498 &        1440 &   33&2   &       0.68 \\
33 & 21:00:06.3 & -33:06:54 & 18.058 & 1.092 & 081397 &        1350 &  -65&2 ::&       0.22 \\
   &            &           &        &       & 102898 &        1800 &  -40&4   &       0.89 \\
34 & 21:00:07.5 & -33:09:56 & 18.040 & 1.032 & 081498 &        1440 &   14&5   &       0.58 \\
\hline
\end{tabular}
\end{center}

\clearpage
{\bf Figure Captions}

\figcaption[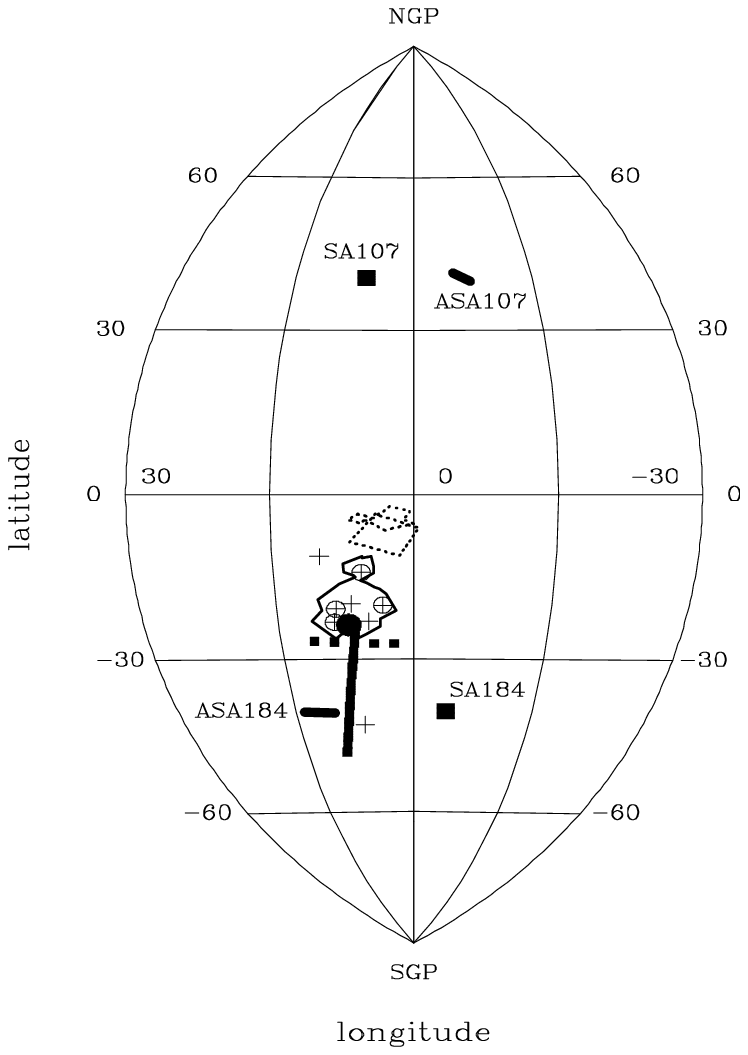]{Aitoff projection showing the location 
of the suspected Sgr globular clusters ({\it circled crosses}), 
previous detections of stars in the Sgr stellar stream 
({\it solid dot} for M96, {\it dotted lines} delineate the 
survey areas of A96 and A97, and {\it solid contour} for the 
approximate outer isophote of I97), Sgr planetary nebulae 
({\it crosses} -- the two {\it crosses} outside the I97 
isophote are only possibly associated), our SA fields 
({\it squares}) and our anti-SA fields ({\it solid strips}).
The long {\it solid line} shows the survey of M98.}

\figcaption[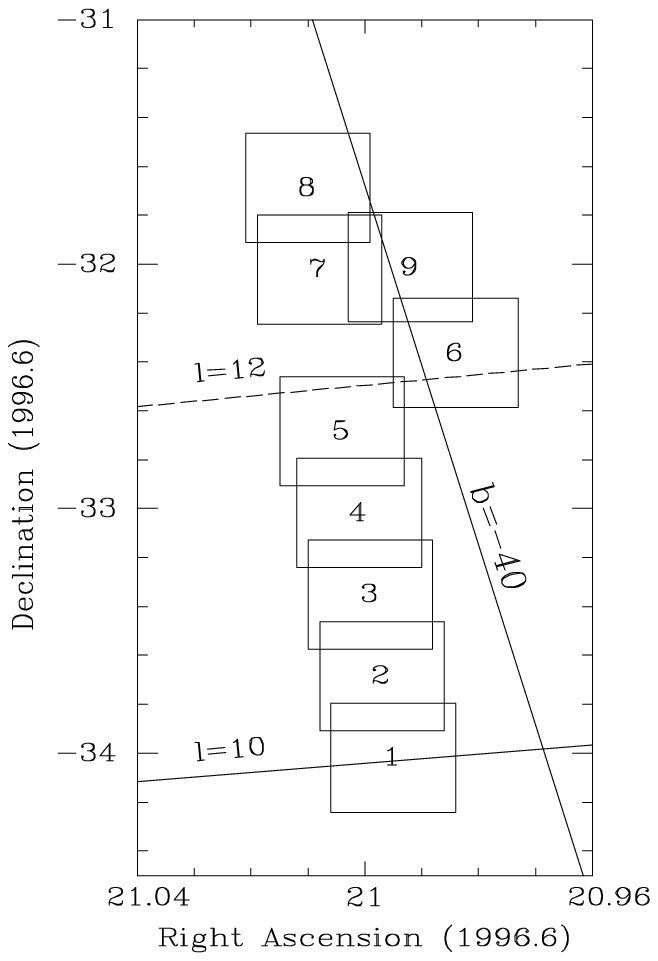]{Placement of the CCD fields in the 
ASA184 strip.  Subfield 6 was displaced to avoid a bright star 
and subfield 9 is intended to provide a check on the 
photometric tie-in between subfields 6 and 7. }

\figcaption[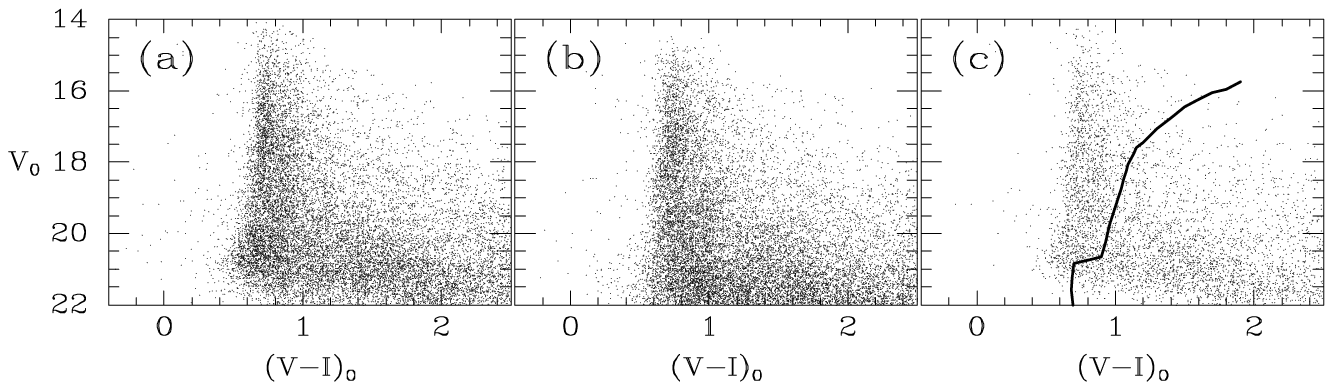]{Dereddened color-magnitude 
diagrams (CMDs) of the (a) ASA184 and (b) ASA107 fields.  
Panel (c) shows the CMD of subfields 2, 3 and 4 of ASA184.  
Although the Sgr red clump may be present in the field, the 
overwhelming contribution of the field star population 
makes it difficult to discern.  A Sgr isochrone from SL95 
is superimposed on top of the data in panel (c).}

\figcaption[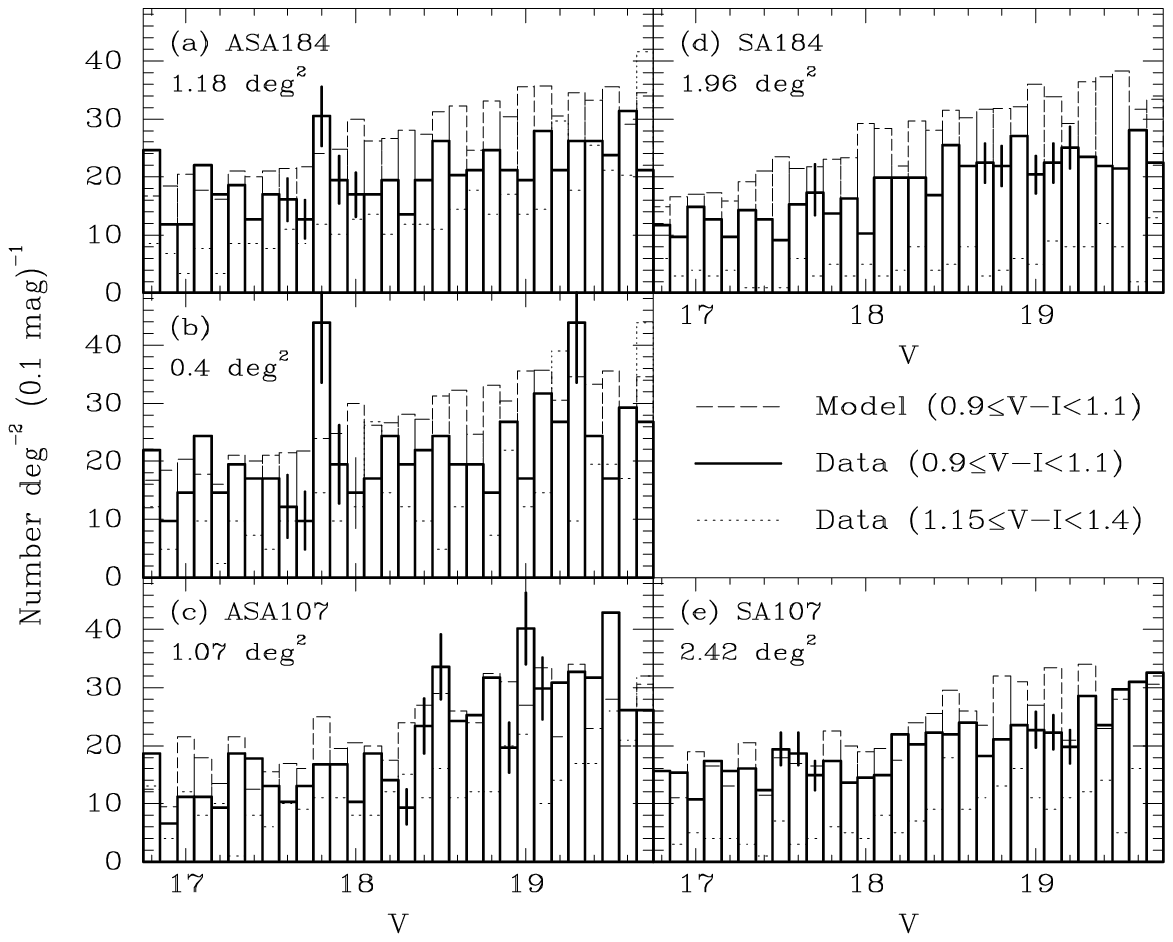]{Magnitude histograms of the 
(a) ASA184 , (c) ASA107 , (d) SA184  and (e) SA107 
fields in the dereddened color interval $0.90 < (V-I)_o < 1.10$.  
In each panel, the {\it dark solid lines} are our 
data normalized to 1 deg$^{-2}$ for the color range 
$0.9 \le (V-I)_o < 1.1$, and, for comparison, our computer 
model predictions for this color range ({\it dashed 
lines}) and the data in a redder color bin 
($1.15 \le (V-I)_o < 1.4$, {\it dotted lines}). 
A few representative error bars are given in each panel.
We have determined $E_{B-V}$ reddenings for each field as
0.08 (ASA184), 0.05 (ASA107), 0.02 (SA184), and 0.07 (SA107).  
Note the weak $V=17.8$ peak in ASA184.  
Panel (b) narrows ASA184 to the 2, 3 and 4 
subfields.  The peak at $V_o=17.8$ is much clearer in this 
smaller ASA184 sample.}

\figcaption[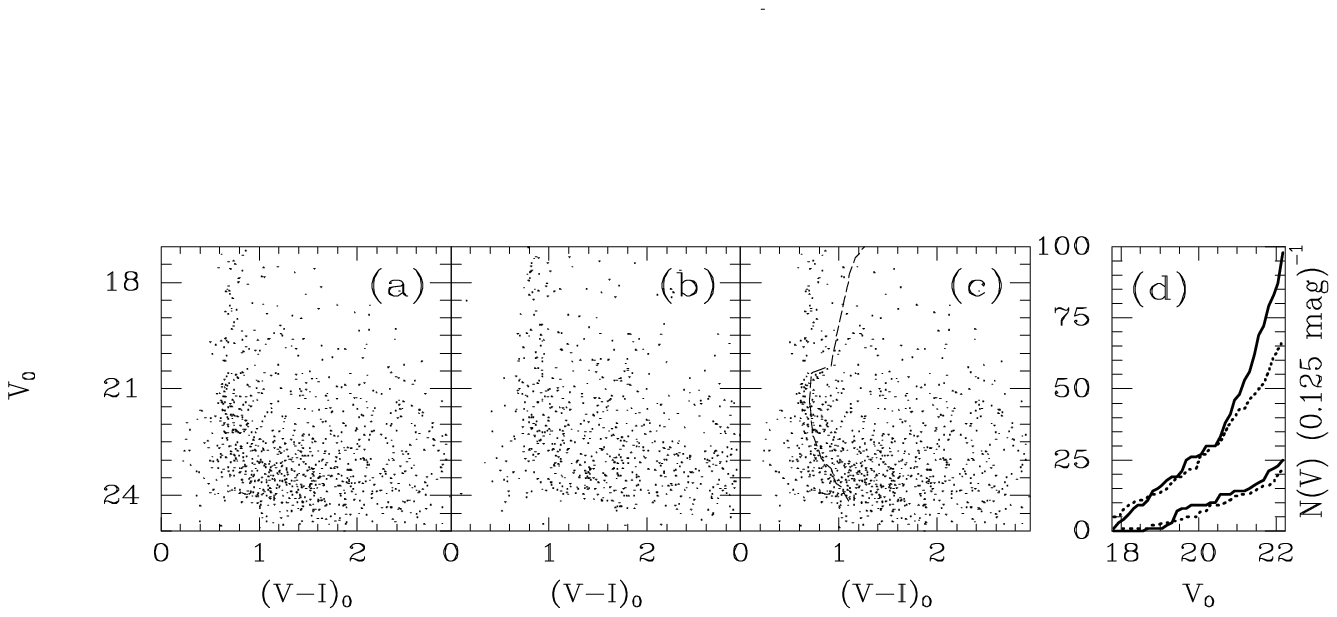]{Dereddened color-magnitude diagrams 
of (a) the ASA184-3 subfield (($l, b$) = ($10.9^{\circ}, -40.2^{\circ}$)), 
which shows the strongest Sgr detection and 
(b) the ASA184-8 subfield (($l, b$) = ($13.0^{\circ}, -40.1^{\circ}$)), 
which is nearly two degrees west (approximately along $b=-40^{\circ}$),
away from the detection.  Panel (c) shows the
ASA184-3 CMD with the F96 isochrone superimposed.  
Panel (d) shows the cumulative starcount distributions 
from the deeper du Pont CCD data for ASA184-3 ({\it solid 
lines}) and ASA184-8 ({\it dashed curves}) for two color ranges: 
$0.6 < (V-I)_o < 0.8$ ({\it upper curves}), which should 
include the bulk of the Sgr main sequence turn off, and 
$1.1 < (V-I)_o < 1.3$ ({\it lower curves}), which should not 
contain many Sgr stars in the magnitude range shown ($17.75 < 
V_o < 22.25$).  The discrepancy between the shapes of the two 
upper curves suggests the presence of
the Sgr MSTO in ASA184-3 (see text). }

\figcaption[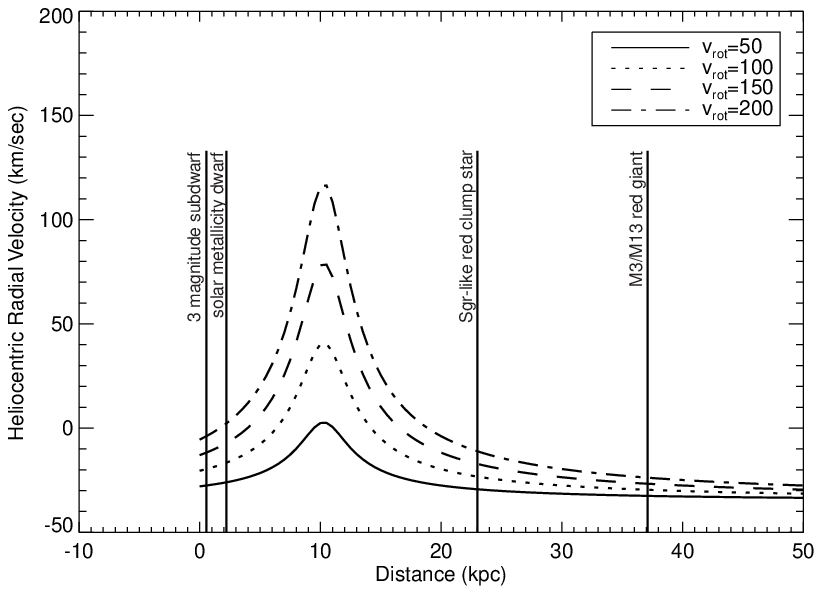]{Expected mean heliocentric radial 
velocities for stars of various mean rotational velocities about
the Galactic center as a function of distance (kpc) in the direction
of ASA184.  The curves are drawn assuming
a mean rotation for the local standard of rest ($\Theta_o$) of 220 km s$^{-1}$, a 
solar motion of $(U,V,W) = (-9,11,6)$ km s$^{-1}$, and a 
solar Galactocentric distance of 8.0 kpc.  For stars of $V_o=17.85$
and $(V-I)_o=1.0$ (the magnitude and color of the ``Sgr red clump"
sample), we mark the approximate distances of stars of different
possible luminosities: solar abundance dwarf (absolute
magnitude $M_V=6.16$ adopted from Reid \& Gizas 1997), a 3.0 magnitude
subluminous subdwarf ($M_V$ as suggested by the subdwarfs in 
Monet et al. 1992), a red clump star of similar luminosity to 
those in Sgr (adopting the I97 $M_V=1.04$), 
and a red giant star of the abundance ([Fe/H] $\sim -1.55$) of
M3 and M13 (absolute magnitude of $M_V \sim 0$ derived from the 
data in Johnson \& Bolte 1998).  The effect of lowering $\Theta_o$
is to offset the curves systematically in the positive direction;
e.g., the Olling \& Merrifield (1998) recommended $\Theta_o = 184$
km s$^{-1}$ increases the expected radial velocities shown by 5 
km s$^{-1}$.
}

\figcaption[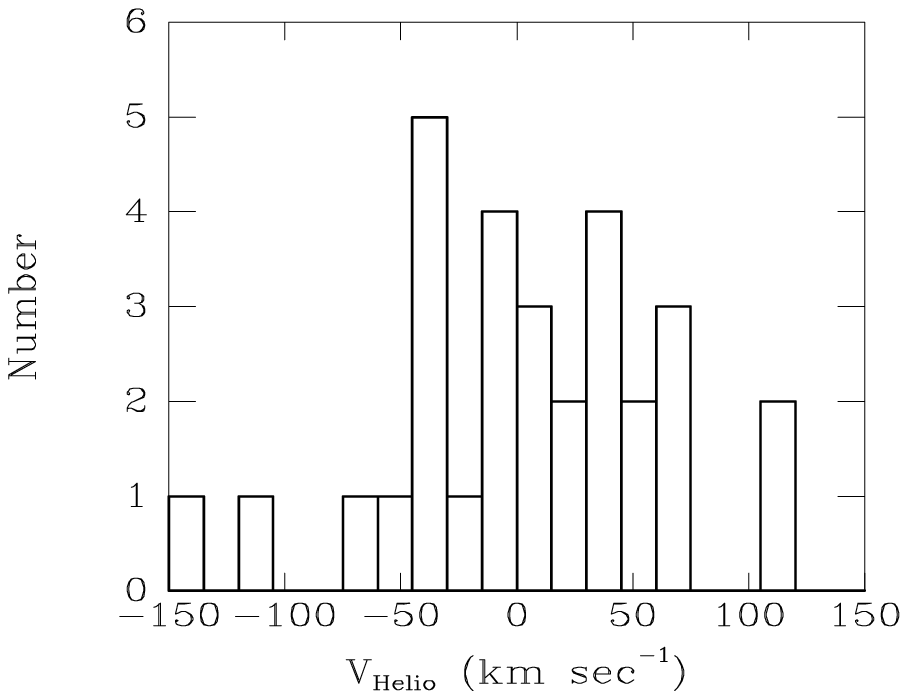]{Histogram of radial velocities for the 
Sgr red clump sample.}

\figcaption[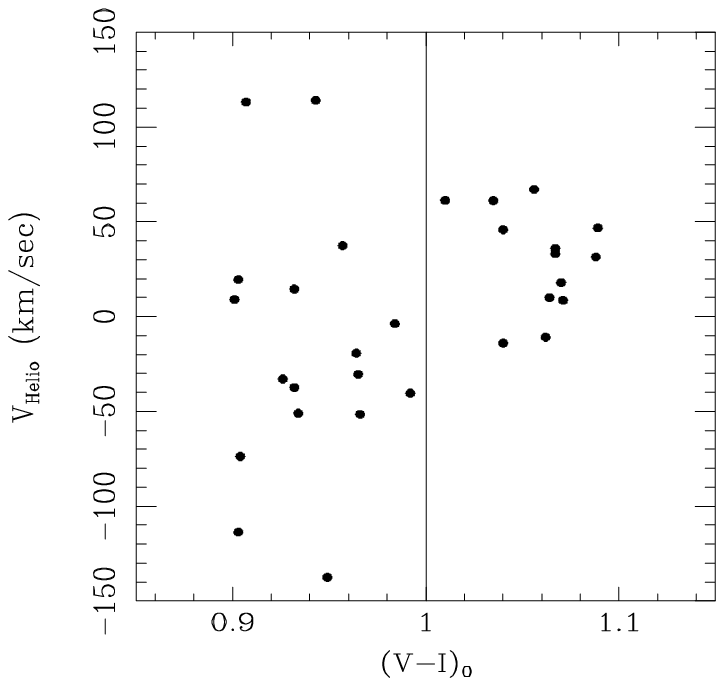]{The distribution of stars in radial velocity
and color.  Note the skew of the redder stars in this sample.  While
it is tempting to assign these stars to Sgr, this assignment
does not necessarily square with the distribution of the count excess
(see text).}

\begin{figure}
\plotone{Majewski.fig1.eps}
\end{figure}
\begin{figure}
\plotone{Majewski.fig2.eps}
\end{figure}
\begin{figure}
\plotone{Majewski.fig3.eps}
\end{figure}
\begin{figure}
\plotone{Majewski.fig4.eps}
\end{figure}
\begin{figure}
\plotone{Majewski.fig5.eps}
\end{figure}
\begin{figure}
\plotone{Majewski.fig6.eps}
\end{figure}
\begin{figure}
\plotone{Majewski.fig7.eps}
\end{figure}
\begin{figure}
\plotone{Majewski.fig8.eps}
\end{figure}

\end{document}